\documentclass{article}

\usepackage{PRIMEarxiv}
\usepackage{amsmath}
\usepackage{amsfonts}
\usepackage[utf8]{inputenc} 
\usepackage[T1]{fontenc}    
\usepackage{hyperref}       
\usepackage{url}            
\usepackage{booktabs}       
\usepackage{amsfonts}       
\usepackage{nicefrac}       
\usepackage{microtype}      
\usepackage{lipsum}
\usepackage{fancyhdr}       
\usepackage{graphicx}       
\graphicspath{{media/}}     

\title{Time Transitive Functions for Zero Knowledge Proofs
}

\author{
  Author1\\
  Ekleen Kaur\\
  University of Florida \\
  \texttt{ekleenkaur17@gmail.com} \\
   \And
  Author2 \\
  Gokul Alex\\
  EPIC Engineering Ventures \\
  \texttt{gokulalexoffice@gmail.com} \\
}
\usepackage{algorithm}
\usepackage{algpseudocode}

\begin{document}
\maketitle

\begin{abstract}
Verifiable delay functions have found a lot of applications in blockchain technology in recent times. Continuous verifiable delay functions are an improvement over the basic notion of VDFs with recursive capabilities.
We are proposing the application of VDF for constructing more space time-efficient provers and simulators required for the iterative non-interactive zero-knowledge systems.
\end{abstract}


\section{Introduction}
Zero-Knowledge proof systems have incentivized its importance in stimulating blockchain security with its mathematical proof. The applicability of zero-knowledge proof in the modern blockchain is enhancing trust amidst its entrants with verifiable proof without revealing the data itself. The proof requires multiexponential prover time in linear verification time for protocols such as bulletproofs, the advantage of bulletproofs is its short cryptographic proof size.  Snarks is another blockchain protocol that escalates verification time due to its short proof size with an initially required trusted setup that ensures the verifier of the veracity of the prover. 

For this paper, the protocol chosen for establishing a correlation between zero-knowledge proof and VDFs is zk-snarks for the reason that the initial trusted setup required in Snarks gives it acceleration over bulletproofs and helps achieve much lesser verification time. Snarks efficiently establish successful entrant verification using Zokrates, a toolbox for deploying zero-knowledge proof verification on Ethereum. 

The definition of probabilistic zero-knowledge digital proof extends to the domain of blockchain transaction data such that the prover(P) can individually verify and validate the transaction data with a probability of providing a shred of cogent evidence to the verifier, to be at least 1-$\epsilon$ (Completeness) and at most $\epsilon$(Soundness). $^{[1][2]}$ We use this verifiable non-interactive probabilistic proof along with deterministic recursive delay functions for offering real-time applicability with a Nash equilibrium for providing security against arbitrary attacks by entrants.

In a proof system with a large susceptible amount of processors running in parallel, the security of the system is contingent upon the inability of the adversary to be able to process the output of the function in arbitrary circumstances in any less than (t) sequential steps for a given verifiable delay (t). VDF’s are sequential, differentiating from an honest prover evaluating the input in (t) steps. So, given with a parallel polynomial no of processing machinery to an adversary, a verifiable delay function makes it impossible to distinguish the function output y corresponding to an input x from random data for any adversary attempting a clandestine attack in significantly fewer steps.
 
VDF’s are unique and verifiable, a VDF consists of a verifier $Setup( \lambda, t )$ → $pp$, for a security parameter ($\lambda$), and a delay of t resulting in a public parameter pp. This public parameter is required for evaluation of input x, $Eval(x, pp)$→  $(y, \pi)$, with the correct output y corresponding to the input x, and a proof $\pi$ required for the verification, $Verify(pp, y, x, \pi)$. This successive order of the protocol makes it efficiently verifiable. VDF also states this constraint that for a given output y on x that is verifiable, it is not possible to hold a second value different from the output y in the preceding evaluation with respect to the same input x, uniqueness.$^{[3]}$

By architecture designing, evaluation of the input x, runs in parallel time $\sigma$(t), on polynomial processors p(t), for the given delay t. Sequentiality $\sigma$(t) of the system in terms of security becomes vulnerable for the parallel processors, as the ideal sequentiality $\sigma$(t) is achieved at, $\sigma$(t) = t -1, which can be taken as $\sigma$(t) =  t - $\epsilon$t, where system security becomes more vulnerable as $\epsilon$→ 1. 

For achieving sequentiality $\sigma$(t) in order to build a VDF, there are approaches, which result in significant as well as insignificant outcomes. In the likelihood of possibly verifying the outcome of the proof points on the elliptic curve, it violates the constraints of time boundation, Eval must be parallel time in $O(polylog(t))$ processors, similarly, the restrictions on t include that it must be subexponential in the $\lambda$. It is possible to obtain a trivial VDF, by iteratively hashing SHA256, this can possibly build a weak VDF,  given the above constraints. 

For a deterministic iterative sequentially verifiable function, where $g: X$ → $X$, is the root function which is sequentially iterated in t steps, such that $f : N × X$ → $X$, then by definition $f ( n, x ) = g^nx$, i.e. iterating the root function n no of times. The mathematical definition of a continuous verifiable delay function, is similar and comprehensive, although the assumption takes into consideration that g is a verifiable delay function with polynomial processing and subsequently iterating n such functions over the same constraints, achieve a CVDF. 

Although for a single VDF such sequentiality trivially achieves a VDF but does not meet the ideal constraints of exponentially faster verification than the prover, despite having soundness. The later section of this research elucidates this mechanism and expounds our original contribution in regard to the incrementally verifiable computation approach, for zero-knowledge proof systems.

\section{Literature Review}

Verifiable Delay functions as discussed in the introduction section of this paper, have a uniqueness constraint in the protocol, lacking this was the work of Mahmoody$^{[4]}$, which stated that the hashing algorithm scheme along with the use of intense robust directed acyclic graphs, will help achieve time lock puzzles in random oracle models, and established the same architecture using Fiat-Shamir non-interactive verification scheme. His model also proposed the gaps that it is impossible to simultaneously use a generator function for solving a puzzle and generating another solution. Mahmoody’s publicly verifiable sequential proof for random oracles using time-lock puzzles is useful against denial of service attacks. Rivest, Shamir, and Wagner$^{[5]}$ assume sequentiality on this time-lock model over exponentiation modulo, the same time-lock concept was applied to the coin flipping scheme by Boneh and Naor$^{[6]}$

Even Cohen and Pietrzak$^{[7]}$ construct a more efficient binary tree with extra edges in place of collision-resistant hashes and DAG for random oracles. Moreover [7] couldn’t cipher the problem of valid outputs for different polynomials, such that for a given solution y, their construction model proved to be valid for some y’, failing the uniqueness constraint. The previously discussed work by [4] postulates the generated proof size to be linear in delay T, i.e. O(N), but [7] reduces it to O(log N) space. 

Meena Jagadeesan’s$^{[8]}$ proof of sequentiality is based on the previously mentioned authors’ contribution, she commensurates, [4]’s construction and [7]’s assumption to bring together a fair possibility of implementing coin flipping scheme as proposed by [6], it takes into account the possibility of a shared random bit, between two entrants and provisions each of a forced commitment in case one entrant aborts, to complete the entire protocol, given it is immune to second preimage attacks. She proposes various improvements on [4]’s drawn conclusions, mitigates the research gap on [7]’s design model and its assumptions.

Lenstra and Wesolowski$^{[9]}$, propose the nearest proof of sequential work (PoSW) by chaining puzzles together into the sloth-hash chain function. They prove a sloth can be verified in a hundred times lesser computation than required to sequentially evaluate it, their work was nearest to a pseudo-VDF. However, still the verification time of their construction couldn’t run in total polylog(t), failing the asymptotic constraint of a VDF.  Benjamin Wesolowski$^{[10]}$ later proposes more efficient verifiable functions by constructing using a group of unknown order elements which result in smaller outputs that are single elements on group functions and streamlined correctness verification. The author posits the potential of these for decentralized networks on the blockchain, random beacons, peer-to-peer trust due to verifiable proofs and resource-efficient distributed applications. However, the paper [9], [10] doesn’t provide any such practical formulation of the posits and does no work in that regard. Our approach is to build such a proof system that requires prover simulators, verifier setup and provides a more secure-trustable system, with both on-chain and off-chain computation, of the authenticated proof and efficiently verifiable delays. 

Proven by Jin Yi Cai$^{[11]}$, the primitive sequential function was supposed to be a finite group of unknown order, repetitively squaring it certain no. of times for iterative sequentiality. Similar to this approach was Burt Kal’s[12] contribution which superseded by stating the utility of hashing functions for this. [12] proposed SHA256 as the hash for achieving this and [4] illustrated it. For these conventions given t iterations on a VDF, successive squaring of this computation along with constant round proofs system heuristics (a variant of Fiat-Shamir protocol) results in a Continuous Verifiable Delay function (CVDF).   
Naomi$^{[13]}$ proposed this from the complexity perspective as this model supersedes the traditional VDF. At any point of the iteration given the state is verifiable in polynomial time $O(polylog$ $t)$ and the system does not require to be reiterated to the designated point, a CVDF is better in terms of solving a harder Nash Equilibrium problem instances that can handle more complex computation which for a simple VDF seems quixotic. 

Consequently, the challenges arising due to more byzantine problem instances are better solvable by CVDF’s as stated by [13], but this research is a computation of its bisimulation with zero-knowledge proof for a single VDF. [4] suggests such verifiable computation for SNARKS achieving $\sigma$(t), where $\sigma(t) = (1 - \epsilon)·t (k+1)·log(t)$ sequentiality, to be nonfunctional. The author conduces to this dichotomy because it is 100,000 times expensive for such a computation on SNARG rather VDF making the approach highly quixotic for the specified constraints and due to this we have adopted the incrementally verifiable approach for achieving $\sigma$(t) which is asymptotically close to the numerator i.e. (1- $\epsilon$)·t and is independent of t. 

Unlike [13] our approach is iterative sequentiality for zero-knowledge proof systems. Incrementally verifiable computation for our model is our novel approach in this research, Boneh$^{[14]}$ mentions the post-quantum security problems for any attack from an adversary with access to a quantum computer. The author’s illustration for just theoretically quantifying the quotients in exponents with attempts of accelerating prover time and the meaningful assimilation of MiMC with STARK for quantum security, addresses some concerns for the traditional VDF’s. However, our setup of a SNARK uses a different approach for addressing some relevant security concerns for zero-knowledge proof applications on blockchain, and through combining these with the incrementally verifiable approach of computation we come up with our model with the aim of eliminating the research gaps.

For an incremental proof of sequential work (IPoS), Dottling$^{[15]}$ proposes provers to individually compute the defined number of steps of the computation to be further carried on by subsequent prover’s from the last computed step without curtailing any prover’s resource. Dottling proposed to achieve this in a 2-type construction, where for the first type dottling had a relatively unique mechanism of considering only a single processor and only relies on hash functions. Mahmoody and Smith$^{[16] [18]}$ propose the construction of a VDF using the black-box construction which explicit ideal hash functions for random oracles. Concurrently Dottling$^{[17]}$ proposes inefficient prover time in Tight Verifiable delay functions (TVDF) by combining the result of non-interactive argument into iterative hashing along side stating the possibility of black-box construction from random oracles to be successfully evaluated in $T + O(T \delta )$ for any constant $\delta$ < 1. Prior to his work, the existence of such TVDF in random oracles was dubious. [18] eliminates these distinguished research gaps for privately and publicly verifiable PoSW.

Unlike [15] that applies a polylogarithmic factor to the proof size, our work has a  subexponential constraint factor in each characteristic step of sequential computation. The characteristic steps for each subsequent iteration enumerate the subexponential factor summing up to the final subexponential running time of the Snark compilation. By doing this we qualify our approach of a Snark delay computable to more intricate problem domains requiring more collective interoperability of proof systems, and as proposed by [13] for more byzantine nash equilibrium problem instances. 

Landerreche$^{[19]}$ unlike our correlation proposes a basic idea of culminating non-interactive cryptographic timestamping and some applications in decentralization, the paper revises on the impossibility of elapsed time in non-interactive timestamping in a UC framework and also limits the time-window for any forged timestamp attack by an adversary. Our correlation with non-interactive cryptography in Snarks and elapsed time-delays for decentralization is for developing better and more secure proof systems.

For establishing circuit verifiability it is required to be adequately tested, so the correctness of timing can be inferred using delay tests$^{[20]}$, such delay tests are applied to the iterative argument of knowledge for arithmetic circuits in our work. Modeling an arithmetic circuit for non-interactive proof using zero-knowledge protocol is a list of manipulation gates in regard to sets of linear consistency equations describing corresponding input and output schemes.$^{[21]}$

These gates are further represented with bilinear consistency equations on those inputs and outputs such that the equation satisfies the arithmetic circuit. These arithmetic circuits are iteratively computed in subexponential iterable time. Benedikt Bunz’s$^{[22]}$ work is on dark np relations, with linear preprocessing requiring succinct communication and logarithmic verification circuit complexity. Unlike his work, our work requires a trusted setup, their work emphasizes transparency, on the other hand we detail security for decentralization concerns; through our approach by analysing our proof system we establish a relatively different evaluation method and a verifiable delay. 

The formulation of the most optimal evaluation scheme from our model, in order to assimilate it with delayed verification for SNARKS is the key motive behind this research. So to summarize, in this section none of the aforementioned research works achieve a much safer decentralised economy against adversaries behind arbitrary attacks in cross domains proof verification systems using a verifiable delay for traditional snarks. To the best of our knowledge this is the first research work that achieves this. The next section expounds the research work and related proofs in detail.

\section{Methodology}

IVC modelling for SNARKS is a three step protocol as in the traditional SNARKS. The non-interactive succinct proof begins with a precomputation instantiated by the verifier. The precomputation is a trusted setup used as a problem instance to be unravelled by the prover, it is a cryptographically encoded parameter scheme which is used to verify the quadratic polynomial points later on the elliptic curve for a multivariate proof.   

Now the reason for a verifiable delay to come into existence for this model is because of our trusted setup, in a trusted setup there may be a conclave amidst the parties, therefore it engenders the need to mitigate the risks encumbered in a trusted setup. 

For a given cryptographic scheme of the prover P with the knowledge of solutions to a polynomial equation over a group of unknown order say H, such that mathematically it convinces a verifier V ( with an arbitrary computational challenge in a trusted setup $\lambda$ using a proof $\pi$ ), for some given input x over the set X corresponding to a unique output y of the polynomial equation over the set Y, we can say 
\begin{center} 
$\forall (x$  $\epsilon$  $X$ \& $y$ $\epsilon$ $Y )$ $\exists$ $\pi$ s.t. $Verify(\lambda,x,y,\pi) = accept$
\end{center}  

As discussed in section 2, building a VDF model for SNARK follows an incrementally iterative approach for Zokrates evaluation scheme(PoSW).  Before delineating into the same we consider an assumption that the computational model in a non-interactive ZKP,  for a language L where P, V, S are taken to be Turing machines where V is a PPT (probabilistic polynomial time$^{[23]}$) Turing machine then, 
\begin{center}
$\forall j$  $\epsilon$  $L$ \& $z$ $\epsilon$ $\{0,1\}^{*} $ $View_{V} [ P(j)$ $\leftrightarrow$ $V(j,z)]$  $\exists$ $S(j,z)$ 
\end{center} 
For even a single interaction $View_{V}$ that happened between P(j) and V(j,z) there exists an efficient Simulator S to reproduce the result from the given input. 
\subsection{Model}
This follows the iterative hashing approach for achieving PoSW. We use SHA256 as the hashing algorithm. Evaluating a SNARK proof using a single processor involves a series of mathematical computations. SNARK has many variants, delineating the QAP variant which uses arithmetic circuits to represent directed acyclic graphs (DAG) are just a representation of the Zokrates proof. 

While evaluating a Zokrates proof, the protocol it follows encodes it into a quadratic equation of a polynomial problem. The significance of the problem is that it holds only when it is instantiated with correct values so that a prover can convince a verifier of the veritability if the equation turns out to be true. While a turing machine does the computation, the range constraints system allows a connection between a proof and a property which uses succinctness by random sampling to ensure an unbiased verifier secret evaluation. Homomorphic encoding scheme is a part of this series of protocols to do the final computation. Our IVC approach for evaluating a VDF for SNARKS follows the traditional convention steps of $Setup, Gen, Eval, Verify$.

\subsubsection{Model Definition}
A VDF $V_{IVC}\rightarrow \{Setup, Gen, Eval, Verify\}$ for Snarks in our model can be encapsulated in four stages as \{TrustedSetup, RandGen, Eval, Verify\} .

$TrustedSetup(\lambda) \rightarrow \{pk,vk\}$: For a given input circuit C:  F$_{L}$ x F$_{R}$ -> F$_{O}$ and output set \{pk,vk\}, the trusted setup generates a publicly verifiable structured reference string for a given linear relation R$_{l}$ over a secret parameter $\lambda$, here F$_{L}$ \& F$_{R}$ are the functions responsible for evaluating the left and right circuit in precomputation to result F$_{O}$. The proving key pk and verifying key vk are later used by P and V in the proof generation and verification phases. The $V_{IVC}$  in this setup encodes input domain X with an output Y which is defined in a relation over the language of Zokrates proof, in our model we use a polynomial equation for the proof as an iterated polynomial permutation.   

$RandGen (G) \rightarrow R$ :  Given a generator G which generates a random number R in a finite range, where R determines the further quantifying iterable steps of computation during evaluation.

$Eval(Hash_{IVC}( P(pk,x,w)), T) \rightarrow (Er, y, \pi)$ :  The SNARK input of proof evaluation are iteratively hashed for a time delay T, the input x, proving key pk and witness w are a Prover function constituent that result in a proof $\pi$ and output y.  $V_{IVC}$  eval hashes(Er) these iteratively computed proofs where corresponding to each step the proof is unique and sequential to the next step. To consider iterative hashing is unique let us take $Hash_{IVC}$ over a function f to be such that there is a possibility of collision, then the mathematical probability of such a case remains negligibly small, it is worth mentioning a multiple iteration count over a 256-bit hash function, on average requires 2$^{128}$ operations which produces a big cycle of average length to be 2$^{128}$, the massive length is large enough to consider iterative hashing to be unique and the possibility of a collision to be improbable in practice.

$V_{IVC}$  establishes this iterative hashing in subexponential time where the deterministic steps of the delay T is a calculable quantified result of some random number R such that Eval is in subexponential time. 

\begin{algorithm}
\caption{ Pre-computation in subexponential time}\label{alg:cap}
\begin{algorithmic}\\
\begin{itemize}
\item For some $R: RandGen(G) \rightarrow R$, the double log function D computes in some random R such that, $D(R) \rightarrow R_{a}$ and exponential of 2 equals logarithmic $(R)^{R_{a}}\rightarrow \alpha $.
\end{itemize}

\While{$\alpha \neq 1 $}\\

\begin{itemize}
     \item  Common input: encode CRS into SRS using non-public randomness, trusted party generates $p_{k}$ , $v_{k}$.
    \item Prover computes random permutation $\pi$, for input $x$ $\epsilon$ $L, \pi \leftarrow (p_{k}, x, w)$ 
    \item Prover sends this $Com(p_{k},x,w)$ to the verifier (has uniform randomness)
    \item Verifier uses this to compute $decom(\pi,v_{k})$; outputs 1 if valid, if not halts and outputs 0. 
\end{itemize}
 
\EndWhile
\end{algorithmic}
\end{algorithm}
For $c= com(x)$ for a random string r, we know that$^{[27]}$ $decom(c)= x$ over $r$. 

$Verify(pk, x,(y, \pi)) \rightarrow \{0,1\}^{*}$ : SNARKS has a Verifier machine that is turing complete and uses proving key pk, input x and proof $\pi$ to output {accept,reject}. Verify for our $V_{IVC}$ model consists of both onchain and offchain computation, each step of the sequentially generated proof is verifiable on chain. So, for T sequential steps which are calculated in $polylog(\lambda)$ queries we have a deterministically calculable delay T.  

This deterministic delay is secure against an adversary with $polylog(\lambda)$ processors$^{[3],[17]}$ as during the computation for evaluating a VDF, SHA256 is parallelizable which allows the adversary to try hundreds of combinations with a quick speed factor q, say q = 100, on a cheap hardware but repeating doesn’t do much here. In our model for the precomputation P such that q > 0. 

Let $t = \lfloor P^{1/q}\rfloor $,  
Then for a given group operation function G,
such that \begin{center} $g_{i}^{P} \epsilon$ G \& $g(x) = \prod g_{i}^{P}(x)$
\end{center}
then,
\begin{equation} \label{eq1}
g(x) = \prod g_{i}^{(t)^{q(q+1)/2}}(x) \; \epsilon \; \text{G}
\end{equation}
\begin{center}
$\forall$ $i = 1,2 … T  $
\end{center}
\begin{center}
$\Rightarrow$ $g_{i}^{t}. g_{i}^{t^{2}}. g_{i}^{t^{3}}… g_{i}^{t^{q(x)}}$ $\epsilon$ G
\end{center}

For any individual $g_{i}^{t^{q}}$ $\epsilon$ G  subexponential in T over the input x, 
we can say  $g_{i}^{t^{(n-3)}}.g_{i}^{t^{(n-2)}}.g_{i}^{t^{(n-1)}}.g_{i}^{t^{(n) (x)}}$ is subexponential in T, so, g(x) in equation 1 is subexponential in T.
    
All the T sequential steps in the function is comparable to solving T challenges on a single turing machine. The model uniqueness is each step of the sequential computation is verifiable which makes it comparable to a CVDF model. So machine P can prove that a certain state is the current state of computation using zero knowledge proof. 

The expression being parallelizable can be computed in parallel by an adversary with a factor of q speed over a normal sequential solver. But this type of attack violates the sequential property of a VDF and  as a first order approximation time taken to solve the challenges increases linearly in q. For practical values of q, this expression is nearly in the order of $q.2^{-256}$; To mount this attack the adversary must compute and solve all the T challenges otherwise it can gain only a factor of two-speed up along with almost negligible probability in $\lambda$. Hence, the scheme is secure for at most T challenges after which new public parameters are required to be generated.  

Entropy of a random variable is the average amount of uncertainty inherent in the variable’s possible outcomes.  Our model uses a factor D which is entropy difference for entering the accept state as the result of publicly generated randomness$^{[24]}$, so for an adversary computing in parallel time $\sigma$(t) and at most $polylog(t)$ processors the probability it can win is nearly negl($\lambda$) as discussed above, irrespective of the probability such a computation is possible without $\pi$ in the case of a VDF$^{[3]}$. 

Our model construction that works on achieving proof of sequential work for zero knowledge proof system, eliminates this gap as only an adversary that can correctly verify $\pi$ on the elliptic curve and obtains the calculable output D which is the result emitted after the verifier successfully validates the proof, can enter the accept state. \\
We can say for a valid D, \\
\begin{equation}
Verify(pk,x,(y,\pi)) \rightarrow D \Rightarrow 1 \tag {Accept}
\end{equation}

Otherwise,

\begin{equation}
    Verify(pk,x,(y,\pi))  \rightarrow Garbage \Rightarrow 0 \tag{Reject}
\end{equation}

\textbf{NP Complete}: This model is said to be np complete over the round function,
\begin{equation}
 g(t) (x) = g \circ g \circ . . . \circ g (x)	 \tag {t times}
\end{equation}
where $g^{(t)}(x)$ is said to be an iterated sequential function, so for $\lambda$ $\epsilon$ N, $g_{\lambda} X$ $\rightarrow X$ $\exists$ ($\epsilon$,t) such that $t = polylog(\lambda)$. $V_{IVC} = \{TrustedSetup, RandGen, Eval, Verify\}$ is said to be in NP as it satisfies the following properties :

\textbf{Completeness}: For a language l in the linear relation $R_{l}$ , where $x$ $\epsilon$ $l$ iff $\exists$ $w$, s.t $R_{l}(x,w) = 1$. Verification algorithm for $polylog(t)$ queries accepts only for $(x, y, \pi)$ which is unique. So for SNARKS, $V_{IVC}$ holds true for, 

\[
Pr \left[
\begin{array}{c|c}
\begin{aligned} Verify(pk, x, (y, \pi)) \rightarrow D = 1 \end{aligned} & \begin{aligned} \{pk,vk\} \leftarrow TrustedSetup (\lambda) \\ R \leftarrow RandGen(G) \\ (Er,y,\pi) \leftarrow Eval(Hash_{IVC}(P(pk,x,w)), T) \end{aligned}\\
\end{array}
\right] = 1
\]

\textbf{Soundness}:  Verification algorithm rejects an adversary A when given $(x , y’, \pi’)$ for any $y \neq y’$ and $\pi \neq \pi’$ over k iterations. So for Snarks, $V_{IVC}$  holds true for,

\[
Pr \left[
\begin{array}{c|c}
\begin{aligned} Verify(pk, x, (y’, \pi’)) \rightarrow D = 1 \\ f(t,x) \neq y \end{aligned} & \begin{aligned} R \leftarrow RandGen(G) \\ (Er,y’,\pi’) \leftarrow A(H(pk,x,w))  \end{aligned}\\
\end{array}
\right] \leq negl(\lambda)
\]

\section{Conclusion and Future Work}
Concluding with the holistic view of analysed outcomes our model is close to a VDF with the current proof system. This research aims to coalesce on-chain and off-chain computation to stimulate the practical applications of $V_{IVC}$, although currently only through SNARKS, however our future works aim to escalate the domain of VDF applications with more random hashes like pederson hash, and with more high level proof systems. 

In the current system, the monolithic process requires indispensable memory control as running the circuit on a singular machine quickly escalates memory bounds, so for a shared memory cluster in a distributed zero knowledge(DIZK)$^{[25]}$ enabling the computation to leverage aggregated cluster memory, we put forward one of the possibilities that our current work aims to extend ahead. Furthermore, Poseidon hash is applicable over the same circuit with 8 times fewer constraints required per message bit over pedersen hash$^{[26]}$, so integrating our existing model using Poseidon Hash is under consideration. We also plan to refine our model with more efficient evaluation mechanisms in our future works.

\section{References}

[1] : Barak Boaz, “ Zero Knowledge Proofs”, 250-262, https://files.boazbarak.org/crypto/lec\_14\_zero\_knowledge.pdf

[2] : B Manuel et al. “Non-Interactive Zero knowledge”, SIAM Journal on Computing 20.6 (1991): 1084-1118. 

[3] : B Dan et al. “Verifiable Delay Functions”, (2019)

[4] : M. Mahmoody, T. Moran, and S. Vadhan. “Publicly verifiable proofs of sequential work.” In Proceedings of the 4th conference on Innovations in Theoretical Computer Science. ACM, 2013.

[5]: Ronald L. Rivest, Adi Shamir, and David A. Wagner, “Time-lock puzzles and timed release crypto”, Tech. Report MIT/LCS/TR-684, MIT, February 1996.

[6]: Dan Boneh and Moni Naor, Timed commitments, CRYPTO (Mihir Bellare, ed.), Lecture Notes in Computer Science, vol. 1880, Springer, 2000, pp. 236–254. 

[7]: Bram Cohen and Krzysztof Pietrzak. “Simple proofs of sequential work.” In Advances in Cryptology - EUROCRYPT 2018 - 37th Annual International 
Conference on the Theory and Applications of Cryptographic Techniques, Tel Aviv, Israel, April 29 - May 3, 2018 Proceedings, Part II, pages 451–467, 2018.

[8]: Jagadeesan Meena et al. “Proof of Sequential Work,” 2018

[9]: A. K. Lenstra and B. Wesolowski. “A random zoo: sloth, unicorn, and trx.” IACR Cryptology ePrint Archive, 2015, 2015.

[10]: Wesolowski B. (2019) “Efficient Verifiable Delay Functions.” In: Ishai Y., Rijmen V. (eds) Advances in Cryptology – EUROCRYPT 2019. EUROCRYPT 2019. Lecture Notes in Computer Science, vol 11478. Springer, Cham

[11]: Jin-yi Cai, Richard J. Lipton, Robert Sedgewick, and Andrew Chi-Chih Yao. “Towards uncheatable benchmarks.” In 8th Structure in Complexity Theory Conference, pages 2–11. IEEE Computer Society, 1993.

[12]: Burt Kaliski. Pkcs \#5: Password-based cryptography specification version 2.0, 2000.

[13]: Ephraim N., Freitag C., Komargodski I., Pass R. (2020) “Continuous Verifiable Delay Functions.” In: Canteaut A., Ishai Y. (eds) Advances in Cryptology – EUROCRYPT 2020. EUROCRYPT 2020. Lecture Notes in Computer Science, vol 12107. Springer, Cham. https://doi.org/10.1007/978-3-030-45727-3\_5

[14]: Boneh, Dan, Benedikt Bünz, and Ben Fisch. "A Survey of Two Verifiable Delay Functions." IACR Cryptol. ePrint Arch. 2018 (2018): 712.

[15]: Döttling, Nico, Russell WF Lai, and Giulio Malavolta. "Incremental Proofs of Sequential Work."

[16]: Mahmoody, Mohammad, Caleb Smith, and David J. Wu. "A Note on the (Im) possibility of Verifiable Delay Functions in the Random Oracle Model." IACR Cryptol. ePrint Arch. 2019 (2019): 663.

[17]: Döttling, Nico, et al. "Tight Verifiable Delay Functions." IACR Cryptol. ePrint Arch. 2019 (2019): 659. 

[18]: Mahmoody, Mohammad, Caleb Smith, and David J. Wu. "Can Verifiable Delay Functions be Based on Random Oracles?." ICALP, 2020.

[19]: Landerreche, Esteban, Marc Stevens, and Christian Schaffner. "Non-interactive cryptographic timestamping based on verifiable delay functions." International Conference on Financial Cryptography and Data Security. Springer, Cham, 2020.

[20]: Ke, Wuudiann, and Premachandran R. Menon. "Synthesis of delay-verifiable combinational circuits." IEEE Transactions on Computers 44.2 (1995): 213-222.

[21]: Bootle J., Cerulli A., Chaidos P., Groth J., Petit C. (2016) “Efficient Zero-Knowledge Arguments for Arithmetic Circuits in the Discrete Log Setting.” In: Fischlin M., Coron JS. (eds) Advances in Cryptology – EUROCRYPT 2016. EUROCRYPT 2016. Lecture Notes in Computer Science, vol 9666. Springer, Berlin, Heidelberg

[22]: Bünz, Benedikt, Ben Fisch, and Alan Szepieniec. "Transparent SNARKs from DARK Compilers."

[23]: Datta A., Derek A., Mitchell J.C., Shmatikov V., Turuani M. (2005) “Probabilistic Polynomial-Time Semantics for a Protocol Security Logic.” In: Caires L., Italiano G.F., Monteiro L., Palamidessi C., Yung M. (eds) Automata, Languages and Programming. ICALP 2005. Lecture Notes in Computer Science, vol 3580. Springer, Berlin, Heidelberg. https://doi.org/10.1007/11523468\_2

[24]: J. Bonneau, J. Clark, and S. Goldfeder. “On bitcoin as a public randomness source.”, https://eprint.iacr.org/2015/1015.pdf

[25]: H. Wu, W. Zheng, A. Chiesa, R. A. Popa, and I. Stoica. “DIZK: A distributed zero knowledge proof system.” In 27th USENIX Security Symposium, USENIX Security 2018, Baltimore, MD, USA, August 15--17, 2018., pages 675--692, 2018.

[26]: Grassi, L., Khovratovich, D., Roy, A., Rechberger, C., Schofnegger, M., “Poseidon: A new hash function for zero-knowledge proof systems.” In: USENIX (2020).

[27]: Lindell, Y., “How to simulate it - a tutorial on the simulation proof technique.” In: Tutorials on the Foundations of Cryptography, pp. 277–346 (2017) 
\end{document}